\documentclass{article}
\usepackage[labelformat=simple]{subcaption}% added<<<<<<<<<<<

\DeclareCaptionLabelFormat{subcaptionlabel}{\normalfont(\textbf{#2}\normalfont)}

\usepackage[english]{babel}

\usepackage[letterpaper,top=2cm,bottom=2cm,left=3cm,right=3cm,marginparwidth=1.75cm]{geometry}

\usepackage{amsmath}
\usepackage{lscape}
\usepackage{graphicx}
\usepackage[colorlinks=true, allcolors=blue]{hyperref}
\usepackage{natbib}
\setcitestyle{round,aysep={},yysep={;}}

\title{\textbf{The Baryonic Mass Estimates of the Milky Way Halo in the form of High Velocity Clouds}}

\author{Noraiz Tahir $^{1,*}$,  Mart\'{i}n L\'{o}pez Corredoira$^{2, 3}$, Francesco De Paolis $^{4, 5, 6}$ \\
$^1$ Department of Physics, School of Natural Sciences (SNS), National University of \\ Sciences and Technology (NUST), H12, 44000, Islamabad, Pakistan.\\
$^2$ Instituto de Astrofisica, de Canarias, La Laguna, E-38205, Teneriefe, Spain.\\
$^3$ Universidad de La Laguna, La Laguna, E-38205, Teneriefe, Spain. \\
$^4$ Department of Mathematics and Physics ``Ennio De Giorgi'', University of Salento, \\ Via per Arnesano, I-73100  Lecce, Italy;\\ 
$^5$ INFN, Sezione di Lecce, Via per Arnesano, I-73100 Lecce, Italy\\
$^6$ INAF, Sezione di Lecce, Via per Arnesano, I-73100 Lecce, Italy  \\
$^*$\textbf{email:}noraiz.tahir@sns.nust.edu.pk
}

\begin{document}
\maketitle

\begin{abstract}
The halo of our Galaxy is populated with a significant number of high-velocity clouds (HVCs) moving with a speed up to $500$ km/s. It is suggested that these HVCs {\it might} contain a non-negligible fraction of the missing baryons. The main aim of the current paper is to estimate the baryonic mass of the Milky Way halo in the form of HVCs in order to constrain a fraction of missing baryons in the form of these clouds. Such findings would give substantial help in the studying halo dynamics of our Galaxy. 

We first estimate the HVCs distance. We consider the most recent and updated HVC catalog, namely the Galactic All Sky Survey (GASS), which, however, covers the southern sky declinations, south of $b \leq 60^\circ$. Following a model presented in the literature, we assume that most of the HVCs (not all of the HVCs in the Milky Way) were ejected from the Magellanic Clouds (MCls) which is at a distance of about $50$ kpc. We assume that the HVCs have a temperature in the range of about $10^2 - 10^4$ K, and are distributed in the Galactic halo as the Navarro-Frenk-White (NFW) profile. Since the GASS survey covers a small portion of the sky, we estimate the number of missing clouds by using Monte Carlo (MC) simulations. The next step will be to estimate the total mass of the Milky Way contained in the form of these HVCs. The total mass resulted to be $\sim (7 \pm 2) \times 10^{9} M_{\odot}$ in the form of HVCs and compact high-velocity clouds (CHVCs).
\end{abstract}


\section{Introduction}
HVCs are interstellar gas clouds that are moving at speeds substantially different (up to several hundred km/s) with respect to the characteristics of the Milky Way rotation. They are primarily composed of neutral hydrogen and are located outside of the Galactic plane. These clouds were discovered in 1963  by \citealt{muller1963} who observed at high Galactic latitudes several diffuse emission structures through the $21$-cm emission lines of HI. Indeed, HVCs are demonstrated by the average line-of-sight projection of the velocity relative to the local standard of rest (LSR), $|v_{LSR}|> 90 - 100~{\rm kms^{-1}}$ \citep{muller1963}. There is a wide range of sizes and shapes of HVCs ranging from large complexes and streams, such as the Magellanic Stream (MS) \citep{mathewson1974magellanic}, or complex C \citep{hulsbosch1968}, to (ultra-)compact and isolated HVCs (CHVCs/UCHVCs) \citep{braun1999, adams2013}. Since their discovery, HVCs have been of great interest, and their kinematics and dynamics have been studied in detail (see Refs. \citealp{wakker1997, woerden2000, smoker2011, wakker2001}), but there are still many ambiguities related to these clouds. 

According to the standard cosmological model ($\Lambda$CDM), the Universe is composed of about $\sim 70\%$ dark energy, $\sim 25\%$ non-baryonic dark matter, and $\sim 5\%$ of baryonic matter \citep{ade2016}. About half of the baryonic dark matter is detected (see Refs. \citealp{burkert2003, cen1999, li2018a, kathine1996, gerhard1996}), but the other half is still undetected at present. This is the so-called ``missing baryon problem''. It has been suggested that a significant amount of these baryons might be present in galactic halos (for details see \citealp{cen2006, fraser2011, nicastro2008, zhang2021, richads2018}), but in what form they are is an open question (see \citealp{tahir2019a, tahir2019b, depaolis1995a, matilla2020}). 

It has also been suggested that HVCs observed in the Galactic neighbourhood are remnants of the formation of the Local Group galaxies and might contain $\sim 85\%$ of the dark matter amount of the Milky Way \citep{blitz1999, martin1999}. HVCs have never been detected in the other galaxy groups except for the Local one \citep{oort1966, zawaan2001, pisano2004}. Also, observations have shown that there are no star formation activities in HVCs \citep{siegel2005, simon2002, hopp2007}. These facts altogether have shown that they reside in galactic halos \citep{westmeier2017}, but, how do HVCs form and evolve? 

It has been hypothesized that gas in the Galactic halo is projected by intergalactic winds, so there is a Galactic Fountain in the halos and disk which causes the formation of HVCs in the halo environment \citep{hulsbosch1973}. In the framework of this hypothesis, various models have been proposed in which the Galaxy is surrounded by a halo of hot gas, where thermal instabilities lead to the formation of warm clouds that rain onto the Galactic disk as HVCs \citep{maller2004, sommer2006, peek2007}. A caution is that these are just toy models and it is not guaranteed that they are correct, since there are many uncertainties on the kinematics and dynamics of the Galactic halo.

It has also been suggested that HVCs have the same kinematics as the Magellanic Stream (MS) and are associated with the MS in the Milky Way \citep{giovanelli1981, mirabel1981}. So, we can say that HVCs are a part of the stream of clouds that are present in the Milky Way \citep{olano2004}. This makes it easier to study HVCs, if we were able to study the process that produces the MS we can get information about the nature and evolution of the HVCs.

The MS consists of gas that is extracted from the Magellanic Cloud (MCl). There are various processes that explain the extraction of this gas (see for details \citealp{meurer1985, moore1994, gardiner1996, yoshizawa2003, connors2006, bekki2007, mastropietro2005}). Using the model of HVCs being a part of the MCls, it is possible to estimate the distances of HVCs  \citep{olano2004, olano2008}. This is the main motivation that leads us to use this assumption later in the paper to estimate the distance of HVCs present in the catalog in \citealt{moss2013}. Starting from this assumption we are able to estimate the total mass of the Milky Way halo in the form of HVCs and CHVCs.

The paper is arranged as follows: in section \ref{distance} we estimate the distance and distribution of HVCs and CHVCs cataloged in \citealp{moss2013}. The distance of these HVCs and CHVCs have been calculated using the model and assumption given by \citealp{olano2004, olano2008}. The HVCs are classified into three major groups: (i) Population MS with sky position $0<l<100$ for $b<-15$ and $180<l<360$ for $-90<b<25$ with radial velocity range from $-400$ km/s to $400$ km/s (ii) Population W with sky position $200<l<320$, $-50<b<50$ with radial velocity range from $90$ km/s to $345$ km/s; and (iii) Population A-C with sky position $0<l<180$ for $b>-15$, \& $100<l<180$ for $b<-15$ with radial velocity range from $-460$ km/s to $150$ km/s. The distances have been calculated by the kinematic (far distance model) (see Section \ref{kinematicmodel}) and dynamic model (near distance model) (see Section \ref{dynamicmodel}). In section \ref{mass} we then estimate the mass of each HVC and CHVC in order to calculate their total contribution to the Galactic halo mass. Since the catalog covers only the southern portion of the sky we need to estimate the distances and masses of the clouds not in the catalogue and then correct the total mass by putting in the mass present in these uncatalogued clouds. To achieve this we used MC simulations and estimated the total number of HVCs and CHVCs that could be observed and then estimated the missing number of HVCs and CHVCs which allows us to estimate the total missing HVC and CHVC mass (see Section \ref{masscorrected}). In section \ref{baryonicfraction} we then estimate the fraction of baryonic mass in the form of these HVCs and CHVCs. Finally, in section \ref{results} we summarize and discuss the obtained results in detail.

\section{Estimating the HVCs distance \label{distance}}
As anticipated, the first step is to estimate the HVC distances. This should lead us to improve our knowledge of the physical properties of these clouds i.e. their mass, density, size and pressure \citep{wakker2001a, hsu2011}. A traditional method to estimate the HVC distance is to use as a reference one or more halo stars, within the angular size of the cloud, and then search for absorption lines with frequencies comparable with the HVC's velocity in the star's spectrum. If the detection is successful, it means that the cloud is in front of the star while if the absorption lines are not detected we can conclude that the HVC is behind the star. In the later case, deeper observations are needed, because the line of sight to the star may have a lower HI column density than that measured by the observations using $9'$ versus a $35'$ beam \citep{wakker2001a}. Using this traditional method halo stars detected with the Sloan Digital Sky Survey (SDSS), or some other surveys enabled astronomers to find the distances of many Complexes \citep{thom2006, wakker2007a, putman2012}.

It has been estimated that HVCs are at a distance $>15$ kpc with respect to the Earth \citep{lehner2022}. In the case of intermediate velocity clouds (IVCs) with $|v_{LSR}|< 100$ km/s these clouds are a distance of about $1-2$ kpc, in the case of HVCs with $|v_{LSR}|< 100-170$ km/s, they are at a distance of about $3-10$ kpc, and in the case of very high-velocity clouds (VHVCs) with $|v_{LSR}|>170$ km/s, they are at a distance of about $>15$ kpc. It has also been detected that there are several CHVCs with $|v_{LSR}|$ that are present at a distance of about $\leq 50$ kpc (see Ref. \citealp{faridani2014} for details).

Other more indirect distance measurements of HVCs are model dependent and include: (i) H$\alpha$ line observations, assuming that the H$\alpha$ emission is the result of the ionizing radiation from the Milky Way galaxy reaching the surface of the cloud \citep{bland1998, bland2002, bland2001, putman2003}; (ii) distance estimates of HVCs under the assumption that HVCs surround the Galaxy, forming a metacloud of $\sim 300$ kpc in size, having mass $\approx 3.0 \times 10^9~M_{\odot}$. In this model, HVCs are assumed to be the remnant of a powerful ``superwind'' that occurred in MCl about $570$ Myr ago as a consequence of the interactions of the LMC and SMC. As result, it is assumed that these HVCs might be magnetic bubbles of semi-ionized gas blown away from the MCls around $570$ Myr ago, and now circulate largely through the halo of the Galaxy as a stream or flow of gas \citep{olano2004}.

\subsection{Kinematic Model \label{kinematicmodel}}
The model is a simplified analysis of the kinematics of the system of HVCs which might help us to solve the problem of obtaining the distance to an HVC from knowing solely its sky position and radial velocity.

Our Galaxy is surrounded by a large cloud of HVCs centered on the MCs on the Galactic position $(l_c, b_c)= (280.5^\circ, -33.9^\circ)$ at a distance from the Sun $d_c= 50.1$ kpc. This metacloud is translating as a whole with a barycenter velocity similar to the spatial velocity of the MCs, since the HVCs were launched from this moving platform, and expanding from this center due to the original velocities with which the HVCs were ejected from the Clouds. By means of a simplified model of the kinematics of the system of HVCs, \citealp{olano2008} derived an analytic formula relating the radial velocity of a HVC to its distance.

Following \citealp{olano2008} we assume that this model is reliable for HVCs and CHVCs categorized as population MS with sky position $0<l<100$ for $b<-15$ and $180<l<360$ for $-90<b<25$ with radial velocity range from $-400$ km/s to $400$ km/s, and population A-C with sky position $0<l<180$ for $b>-15$ with radial velocity range from $-460$ km/s to $150$ km/s.

\begin{figure}
	\centering 
	\includegraphics[width=0.8\textwidth]{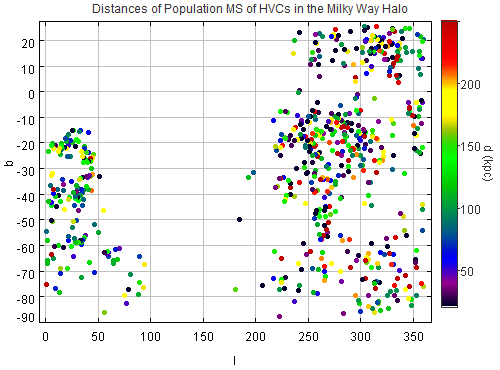}	
	\caption{The distribution of the observed population MS of HVCs by \citealp{moss2013} in the Milky Way halo. The colour scale represents the estimated distance from eq. (\ref{distance}) of each cloud from the Sun. As it is seen that these are distributed in the Galactic halo up to distances of $20$ -- $250$ kpc which is in agreement with \citealp{olano2008}.} 
	\label{fighvcms}%
\end{figure}

The distance of a given HVC, at galactic coordinates $l$ and $b$, moving with radial velocity ${\rm v_{r}}$, can then be estimated by using the relation \cite{olano2008}

\begin{eqnarray}\label{distance}
 & d(l,b,{\rm v_r}) = k^{-1}({\rm v_r} +V_{fall}+ V_s \sin l \cos b+ V_f \\ \nonumber & \cos (l-l_{\circ})  \cos (b-b_{\circ})) + d_c  (\cos b \cos (b_c) \cos (l-l_c) \\ \nonumber & + \sin b \sin(b_c))~{\rm pc},
\end{eqnarray}

where $d$ is the cloud distance in pc. Moreover, $k= (1/550)$ ${\rm km~s^{-1}~pc^{-1}}$, $V_{fall}=133~{\rm km~s^{-1}}$ is a velocity component of the HVC obtained towards the Galactic center due to the action of the gravitational force of the Milky Way, $V_s=220~{\rm km~s^{-1}}$  corresponds to the velocity component of the solar motion in the $(l, b)$ direction of the HVC, $V_f=282~{\rm km~s^{-1}}$ is the translation velocity of the cloud with respect to the Galactic center, and $(l_\circ, b_\circ)=(78^\circ, 2^\circ)$ the direction of the HVC stream. We note that the parameters appearing in eq. (\ref{distance}) with the values given above are the best-fit parameters of the model by \citealp{olano2008}.

\begin{figure}
	\centering 
	\includegraphics[width=0.8\textwidth]{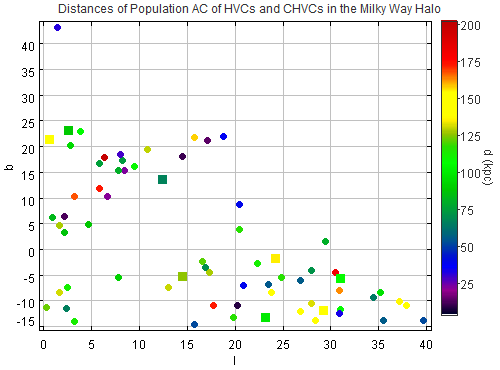}	
	\caption{Same as Fig. \ref{fighvcms}, but for the case of CHVCs.} 
	\label{figchvcms}%
\end{figure}

\begin{figure}
	\centering 
	\includegraphics[width=0.8\textwidth]{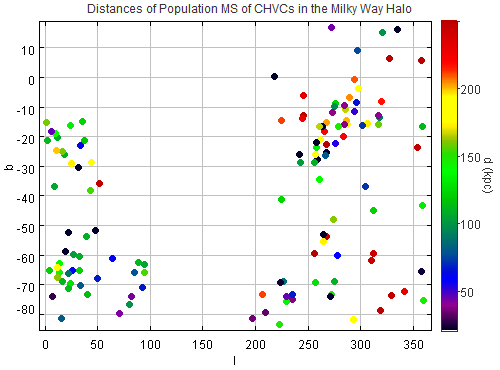}	
	\caption{The distribution of the observed population AC of HVCs (circular points) and CHVCs (boxed points) by \citealp{moss2013} in the Milky Way halo. The estimated distance from eq. (\ref{distance}) is in the range $5$ -- $200$ kpc, which is again in agreement with the results obtained by \citealp{olano2008}.}
  \label{fighvcandchvcfar}
\end{figure}

In Figs. \ref{fighvcms} -- \ref{fighvcandchvcfar} we give the galactic coordinates $(l, b)$ of the observed HVCs and CHVCs by \citealp{moss2013} with their estimated distance (see the color table) through eq. (\ref{distance}) and the data available by \citealp{moss2013}. It is seen that most of the clouds are in the halo of the Milky Way, for population MS of HVCs and CHVCs their distance ranges from $\simeq$ $20$ kpc to about $250$ kpc, and for population AC their distance ranges from $\simeq$ $5$ kpc to about $200$ kpc. These estimated distances are in the distance limits given in Table 1 by \citealp{olano2008}. 
One can also see that HVCs and CHVCs are missing in a wide region of Figs. \ref{fighvcms} -- \ref{fighvcandchvcfar}. This is because only the southern sky hemisphere has been surveyed by the Parkes telescope to produce the catalog by \citealp{moss2013}. We shall estimate the number of missing clouds in the under-populated region, assuming that the missing clouds have the same characteristics as those observed. 

\subsection{Dynamical Model \label{dynamicmodel}}
The model assumed that the HVCs and CHVCs in the Milky Way halo are formed due to the collision of LMC and SMC. There are physical evidences that this collision happened and triggered the processes that gave origin to the HVCs. In their model, the initial conditions of HVCs are found by tracing back the orbit of LMC and SMC which was governed by the Milky Way gravitational potential. With this purpose \citealp{olano2008} estimated the positions and spatial velocities of both LMC and SMC and a law for the force field of the Galactic potential. They considered two different models: \textbf{Model-I}: in which the Galactic potential is due to a dark matter halo; and \textbf{Model-II}: in which the additional potential of a massive disk is taken into account. It was seen from Model-I that the encounter time between LMC and SMC is $850$ Myr, and from Model-II it was $720$ Myr (see \citet{olano2008} for details). It is also stated by \citealp{olano2008} that Model-I is more convenient to use, as it is simpler to apply, and the results from both models coincide with each other. So, for simplification, we will also use Model-I and assume that the HVCs were ejected from the MCs when the Clouds meet at the time $t_e= 850$ Myr ago. 

Following \citealp{olano2008} we assume that the dynamic model is more reliable for HVCs and CHVCs categorized as population W with sky position $200<l<320$, $-50<b<50$ with radial velocity range from $90$ km/s to $345$ km/s, and population A-C with sky position $100<l<180$ for $b<-15$ with radial velocity range from $-460$ km/s to $150$ km/s.

The equation of motion of a HVC subject to the gravitational field of the Galaxy and the LMC, and a drag force $\vec{F}_d$, experienced by the HVC when moving through the gaseous disk of the Galaxy is given by
\begin{equation}
\vec{\ddot{r}}+\vec{F}_h(\vec{r}, r_h)+\vec{F}_{L,HVC}(\vec{r}_{L,HVC})+\vec{F}_d(R, z, v_r)=0,
\label{hvcequationofmotion}
\end{equation}
where, $\vec{r}$ is the position vector whose components are expressed in a Cartesian coordinate system (X, Y, Z), with the origin at the Galactic center, the X-axis pointing in the direction of the Sun's Galactic rotation, the Y-axis pointing in the direction from the Galactic center to the Sun, and the Z-axis pointing toward the Galactic north pole, $F_h(\vec{r}, r_{h})$ is the  Milky Way halo force exerted on the considered HVC and can be written as \citep{olano2008}
\begin{equation}
\vec{F}_h(\vec{r}, r_{h})=\frac{M_h V_c^2 |r_L|}{r^3}\frac{{\rm erf}(r/r_h)}{{\rm erf}(|r_L|/r_h)}\vec{r}.
\label{haloforcelaw}
\end{equation}
Here, $M_h \approx 4.24 \times 10^{11}~M_{\odot}$ is the total Milky Way halo mass, $V_h=220$ km/s is the circular velocity of rotation due to the halo, $\vec{r}_L = (151.96, -31.56, -52.86)$ kpc is the adopted position vector of LMC at $t_e=850$ Myr following Model-I, and $r_h=42$ kpc is the halo scale radius. Also, $\vec{F}_{L, HVC}(\vec{r}_{L,HVC})$ is the gravitational force of the LMC exerted on the HVC and is given by \citep{olano2008}
\begin{equation}
\vec{F}_{L, HVC}(\vec{r}_{L,HVC})= \frac{G~M_L^2}{(|\vec{r}_{L,HVC}|^2+K^2)^{3/2}}\vec{r}_{L,HVC}.
\label{gravitationalforceLMCtoHVC}
\end{equation}
Here $M_{L}=8.7 \times 10^9M_{\odot}$ is the mass of the LMC, $K=3$ kpc is a parameter of Plummer's potential, and $\vec{r}_{L,S}=\vec{r}_L-\vec{r}_{HVC}$  is the position vector between LMC and HVC, and $\vec{r}_{HVC}$ the position vector of the considered HVC respectively. Moreover, the drag force  $\vec{F}_d(\vec{r},\vec{\dot{r}})$ can be given by \citep{olano2004}
\begin{equation}
F_d(R, z, v_r)= \frac{M_d~\sigma(R)}{2h(R)}\exp\left[\frac{-|z|}{h(R)}\right]\frac{v_r}{\overline{N_{HI}}} \vec{v_r},
\label{dragforce}
\end{equation}
where $v_r$ is the radial velocity of the considered HVC $\overline{N_{HI}}$ is the average column density in ${\rm cm^{-2}}$ of the cloud (see the catalog by \citealt{moss2013}), $R=\sqrt{x^2+y^2}$ in kpc, $\sigma(R)= 1.09 \times 10^{21} \exp\left(-\left[\frac{8.24-R}{1.06 \times 8.24}\right]^2\right) \\ {\rm atoms~cm^{-2}}$ is the surface density of interstellar HI, and $h(R)= 105.9 \exp\left(-\left[\frac{3.37-R}{4.23 \times 3.37}\right]^2\right)~ {\rm pc}$ is the scale height of the thickness of the HI gas layer at $R$ \citep{olano2004}. 

We can estimate the dynamical distance of the considered HVC which is given by
\begin{eqnarray}
     & r^2-r_s^2=d^2\left[(\cos b \sin l)^2+(\cos b \cos l)^2+ \sin ^2b\right]\nonumber \\ &-2r_sd\cos b \cos l.
\label{dynamicaldistance}
\end{eqnarray}
Here, $r_s=8.5$ kpc is the Galactocentric distance of SMC with respect to the Sun. To estimate the distances of the considered HVCs and CHVCs we have to solve eq. (\ref{hvcequationofmotion}) numerically with the given boundary condition that at $t= 850$ Myr the HVC lies at a determined distance $d$ from the Sun in the direction $(l,b)$. This solution will give us the present position vector of HVC $r$. 

\begin{figure}
	\centering 
	\includegraphics[width=0.8\textwidth]{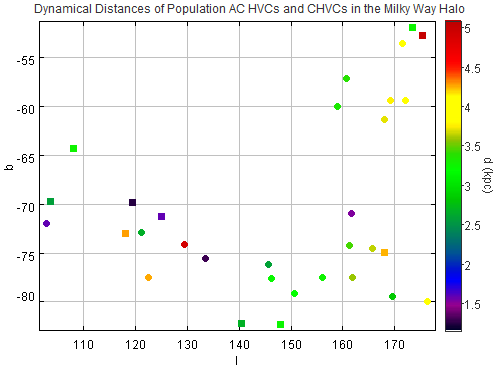}	
	\caption{The distances of the observed HVCs (circular points) and CHVCs (boxed points) with sky positions $100<l<180$ for $b<-15$ and radial velocity range from $-460$ km/s to $150$ km/s in the Milky Way halo. The estimated distance is in the range between $\sim 1$ -- $5$ kpc which agrees with the distance limits by \citealp{olano2008}.}
  \label{fighvcandchvcnear}
\end{figure}

\begin{figure}
	\centering 
	\includegraphics[width=0.8\textwidth]{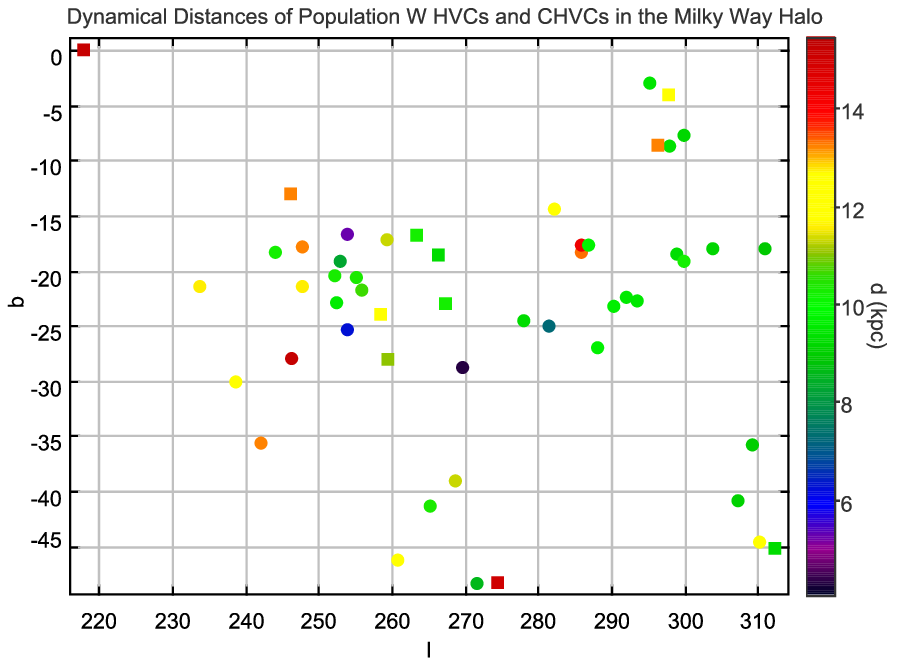}	
	\caption{Same as Fig. \ref{fighvcandchvcnear} but of HVCs and CHVCs with sky position $200<l<320$, $-50<b<50$ and radial velocity range from $90$ km/s to $345$ km/s. The estimated distance is in the range $\sim 5$ -- $15$ kpc which again agrees with the distance limits by \citealp{olano2008}.}
  \label{fighvcandchvcw}
\end{figure}

The distance of the considered HVCs and CHVCs population using the dynamical model is shown in Figs. \ref{fighvcandchvcnear} - \ref{fighvcandchvcw}. 

\section{Total HI Mass of HVCs and CHVCs \label{mass}}
Once the distance of the HVCs and CHVCs has been estimated, one is in a position to estimate the cloud HI mass, since HI is normally dominant in HVCs although ${\rm H_2}$ has been detected. The HI mass can be estimated by using the relation (see Section 3 of \citealt{bruns2001})
\begin{equation}
M(d, \overline{N_{_{HI}}})= \left(\overline{N_{_{HI}}}\times \Sigma \times m_{HI}\right) \left(\frac{d}{R_H}\right)^2.
\label{massequation}
\end{equation}

\begin{figure}
	\centering 
	\includegraphics[width=0.8\textwidth]{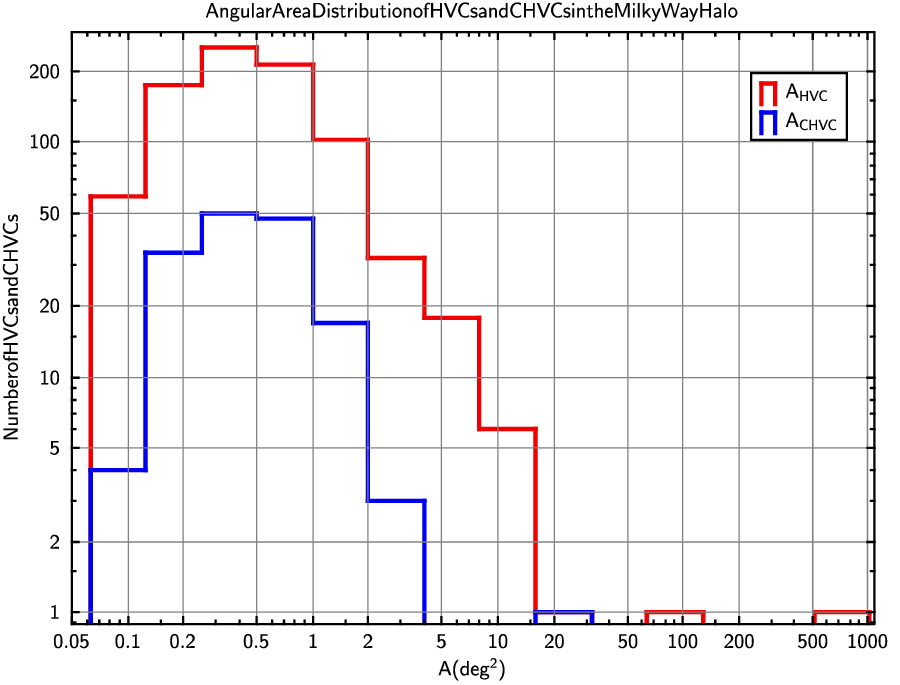}	
	 \caption{The angular area distribution of HVCs and CHVCs in the Milky Way halo. The red and blue curves represent the angular area of HVCs and CHVCs respectively. Note that the selection effects are not taken into account in this plot.}
  \label{angulararea}
\end{figure}

\begin{figure}
	\centering 
	\includegraphics[width=0.8\textwidth]{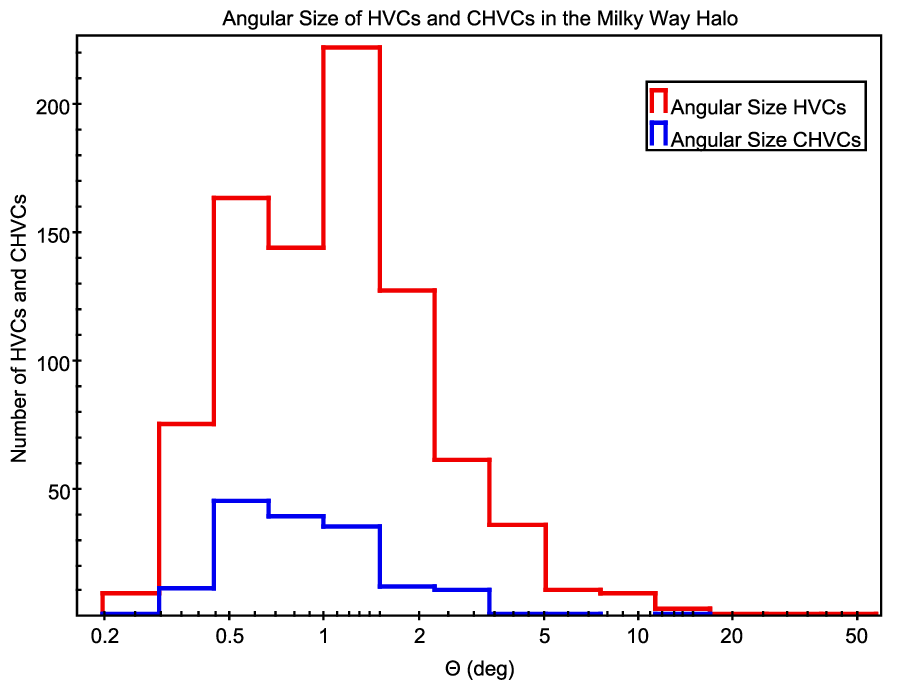}	
	\caption{The angular size distribution of HVCs (red curves) and CHVCs (blue curves) in the Milky Way halo. The data is taken from the catalog by \citealp{moss2013}.}
  \label{angularsize}
\end{figure}

\begin{figure}
	\centering 
	\includegraphics[width=0.8\textwidth]{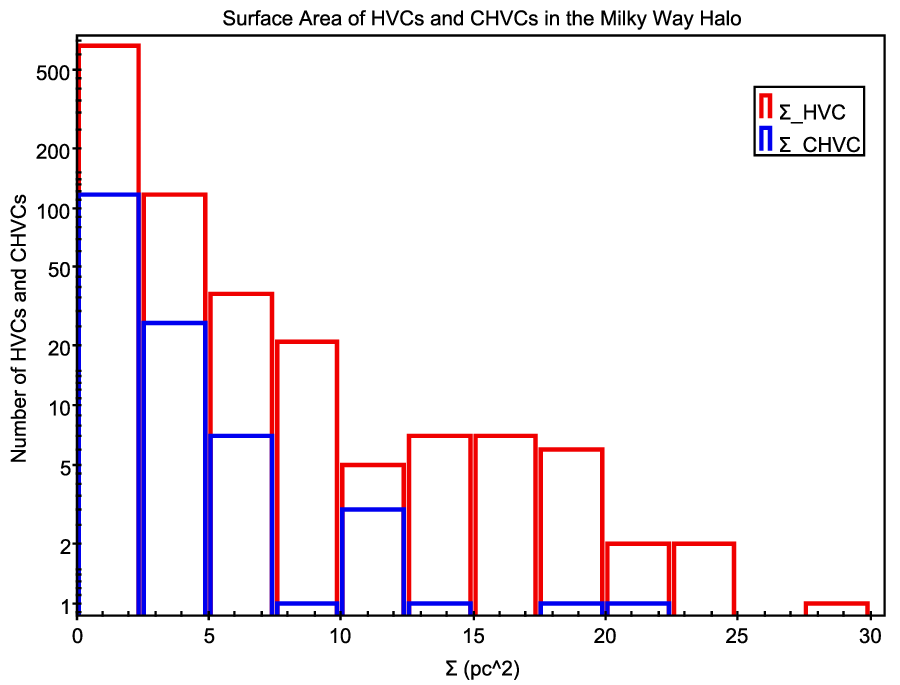}	
	\caption{The surface area distribution of HVCs and CHVCs in the Milky Way halo.}
  \label{surfacearea}
\end{figure}

Here, $m_{HI}= 1.67 \times 10^{-24}$ g, is the mass of a single H atom, $d$ is the distance of the considered HVC or CHVC estimated from eq. (\ref{distance}), $R_H \approx 200$ kpc is the Milky Way halo radius (see \citealt{zaritsky1989}), and $\Sigma$ is the surface area of a single cloud. In order to calculate $\Sigma$ we used the angular area $A$ (see Fig. \ref{angulararea}), the angular size $\theta$ (see Fig. \ref{angularsize}) which are already present in the catalog (see \citealt{moss2013}), and the cloud distance $d$ calculated from eq. (\ref{distance}). In Fig. \ref{surfacearea} we give the surface area distribution of HVCs and CHVCs in the Milky Way halo.

\begin{figure}
	\centering 
	\includegraphics[width=0.8\textwidth]{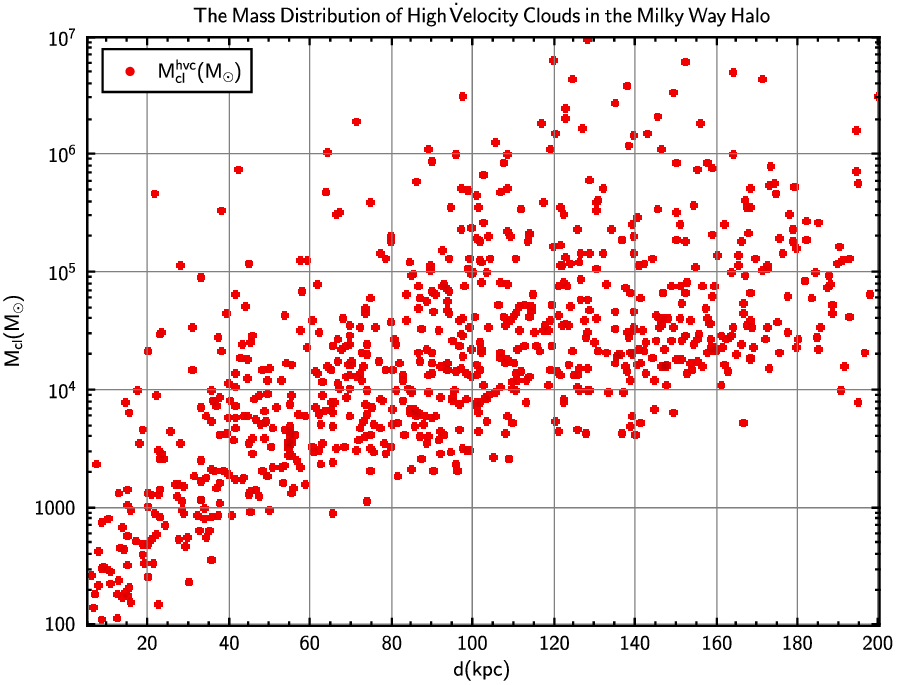}	
	\caption{The estimated mass distribution of the observed HVCs in the Milky Way halo up to $200$ kpc. The total number of observed clouds is $784$, and the estimated total mass assuming distance predicted from eq. (\ref{distance}) is $\approx (6 \pm 1) \times 10^9 \, M_\odot$.}
  \label{fig2a}
\end{figure}

\begin{figure}
	\centering 
	\includegraphics[width=0.8\textwidth]{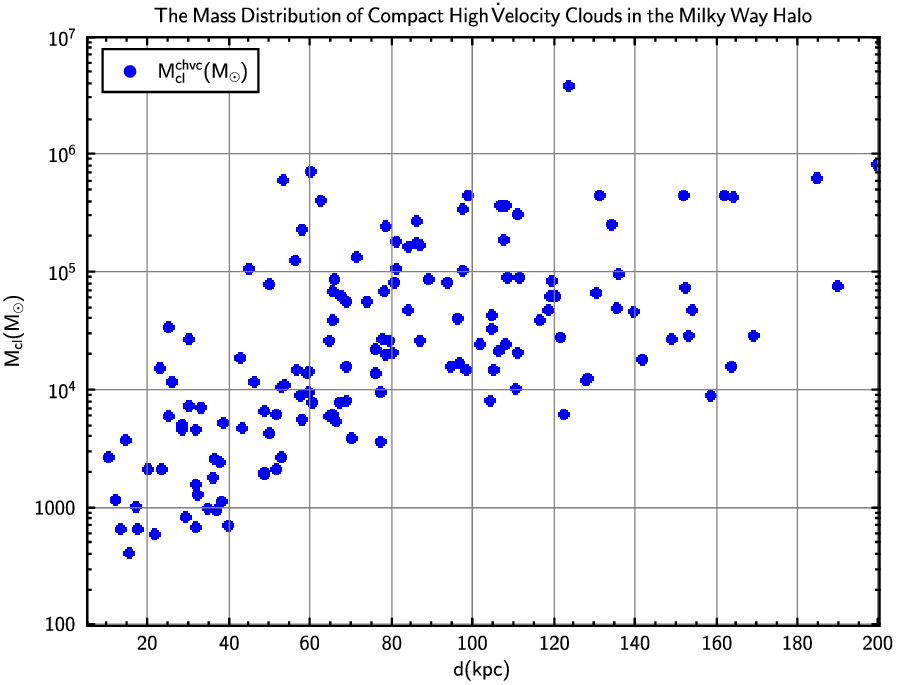}	
	\caption{Same as Fig. \ref{fig2a}, but for the case of the observed CHVCs. The total number of observed clouds is $148$, and the estimated total mass is $\approx (3 \pm 2) \times 10^7 \, M_\odot$.}
  \label{fig2b}
\end{figure}

Using eq. (\ref{massequation}) we can then estimate the mass of each observed HVC and CHVC, which is given in Figs. \ref{fig2a} and \ref{fig2b}.

In order to check the consistency of the cloud mass estimate we need to get an estimate of the maximum mass of an HVC or CHVC. The maximum mass of a molecular cloud is known as the Bonnor-Ebert Mass, which can be written as \citep{bonnor1956, ebert1955}
\begin{equation}
M_{max}(T, P)= 1.18 \left(\frac{k_B T}{\mu m_{HI}}\right)^2 \left(\frac{1}{P^{1/2}G^{3/2}}\right),
\label{bonnorebertmass}
\end{equation} 
Here, $k_B$ is the Boltzmann constant, $\mu$ is the molecular weight, $G$ is Newton's gravitational constant, and $T$ is the temperature of the considered cloud which is $\approx 500$ K which is explicitly given in the catalog by \citealp{moss2013}. Also, $P$ is the gas pressure of the medium. Here, we can write $P=\overline{N_{HI}}k_B T/1.18~R_H$, which implies that eq. (\ref{bonnorebertmass}) can be written as 
\begin{equation}
M_{max}(T)= \left(\frac{1}{m_{HI}^2}\sqrt{\frac{R_H}{1.18 \overline{N_{HI}}}}\right)\left(\frac{k_B T}{G}\right)^{3/2}.
\label{manximummass}
\end{equation} 
Here, $\overline{N_{HI}} \approx 10^{18}~{\rm cm^{-2}}$ is the average column density of the HVCs, and CHVCs \citep{moss2013}. Substituting all the values in eq. (\ref{manximummass}), the maximum mass of the HVCs and CHVCs should be $\approx 10^{9} M_{\odot}$. However, as we can see from Figs. \ref{fig2a}-\ref{fig2b}, the estimated mass of the HVCs and CHVCs is always $M_{cl}\leq 10^{9} M_{\odot}$, in agreement with the expectations. We would like to caution that here we have used spherical symmetry of HVCs, just like for molecular clouds. If we assume fragmented structure or distorted symmetry the maximum mass of HVCs does not change much.
\section{Estimating the Total Missing Mass in HVCs and CHVCs \label{masscorrected}}
As discussed in the previous section we have been able to estimate the distance and mass of the HVCs and CHVCs present in the catalog by \citealp{moss2013}. It is clearly seen from Figs. \ref{fig2a}-\ref{fig2b} that there are missing clouds in the Milky Way halo. In fact, beyond $\sim 40$ kpc we do not find HVCs with a mass smaller than $\simeq 10^4~M_{\odot}$, and the same also holds for the case of CHVCs. However, selection effects are likely important and one can think that while large mass clouds are generally detectable lower mass clouds can be detected only if they are sufficiently close to the Earth. It is therefore important to estimate the number and mass of the missing clouds since they might give a non-negligible effect in estimating the total mass in the form of HVCs and CHVCs.

In order to estimate the missing HVCs and CHVCs we performed a MC simulation to estimate the total number of HVCs and CHVCs in the Milky Way halo up to $250$ kpc. The MC simulation is a computational technique used to model and simulate complex systems or processes that involve randomness or uncertainty. It involves the generation of a large number of random samples based on specific probability distributions and using these samples to estimate or approximate the behaviour or outcomes of the system. 

In our case, the HVCs and CHVCs are assumed to be contributing to the dark matter in the Milky Way halo, and therefore we assume that their distribution might follow the dark matter distribution profile, and in particular the Navarro-Frenk-White (NFW) model which is given as \citep{navarro1996}
\begin{equation}
\rho(r)= \frac{\rho_{\circ}}{(r/r_{\circ})(1+(r/r_{\circ}))^2},
\label{NFWprofile}
\end{equation}
where $\rho_{\circ}= 1 \times 10^{-4}~M_{\odot}/kpc^3$, and $r_{\circ}= 8$ kpc, are the core density and radius of the Milky Way halo \citep{nesti2013}.

\begin{figure}
	\centering 
	\includegraphics[width=0.8\textwidth]{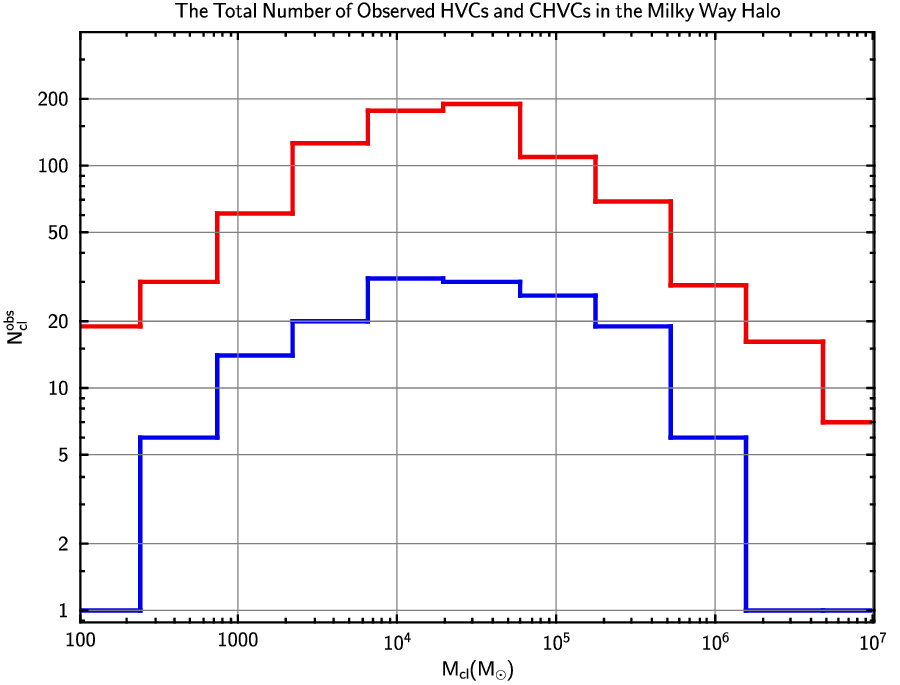}	
	\caption{The total number of HVCs (red line) and CHVCs (blue line) is given as a function of the cloud mass. It is seen that the value of the power law index $\alpha$ for HVCs is $-0.46$, and for CHVCs it is $-0.63$ (see text for details). We can also see that the minimum mass for both these cases is $\sim 10^2M_\odot$ and the maximum mass is $\sim 10^7M_\odot$.}
  \label{fig:numberestimatedhvcandchvcv}
\end{figure}

Once the distribution of the clouds in the halo is assumed, in order to perform MC simulation we have to derive a probability density function (PDF). The expression for the PDF can be derived by using the mass function (MF). To obtain the MF from the NFW distribution we assume that the mass within a given $r$ follows a power-law relationship and is generally given as  \citep{salpeter1955}
\begin{equation}
\zeta(M_{cl}(r))= AM_{cl}(r)^{-\alpha},
\label{eq:mfgeneralhvc}
\end{equation}
where $M_{cl}(r)$ is the mass of the clouds within a radius $r$, $A$ is the normalization constant. The power-law index $\alpha$ can be estimated by the relation
\begin{equation}
\alpha= -\frac{\ln N_{cl}^{right}}{\ln M_{cl}^{total}},
\label{eq:alpha}
\end{equation}
where, $N_{cl}^{right}$ is the total number of observed HVCs or CHVCs whose mass is $\geq 10^4 ~ M_\odot$, and $M_{cl}^{total}$ is the total mass of the observed HVCs or CHVCs. It can be seen from Fig. \ref{fig:numberestimatedhvcandchvcv} that the value of, $\alpha$, is $-0.46$ for the case of HVCs, and $-0.63$ for the case of CHVCs. The values of $\alpha$ are estimated only for the case of the clouds with masses $\geq 10^4\, M_\odot$. If we consider a case for all the clouds $\alpha$ increases slightly for both cases which impacts the overall results of the simulation negligibly.
The total mass of the clouds $M(r)$ within $r$ can be defined as
\begin{equation}
M_{cl}(r)= \int_{0}^{R_{halo}} 4 \pi r^2 \rho(r) dr,
\label{eq:totalmass}
\end{equation}
where $R_{halo}$ is the size of the considered halo. From eq. (\ref{eq:mfgeneralhvc}) we can define the PDF as
\begin{equation}
P(M_{cl}(r))= \frac{\zeta(M_{cl}(r))}{\int_{M_{min}}^{M_{max}} \zeta(M_{cl}(r)) dM_{cl}(r)},
\label{eq:pdf}
\end{equation}
where $M_{min}$ is the minimum, and $M_{max}$ is the maximum estimated mass of the HVCs and CHVCs, respectively. In order to estimate $A$ we use the normalization condition for MF, which is given as \citep{kroupa2001}
\begin{equation}
\int \zeta(M_{cl}(r)) dM_{cl}(r) = 1.
\label{eq:normalization}
\end{equation} 
We are now in a position to perform MC simulation. It is clearly seen from Fig. \ref{fig:numberestimatedhvcandchvcv} that there are $784$ HVCs, and $147$ CHVCs present in the catalog. In total, we have about $932$ clouds observed in the Milky Way halo within $200$ kpc. We perform MC simulation for HVCs and CHVCs separately. As we know the total number of these clouds, $N_{total}$, we run the simulation and estimate the probability of finding a HVC and CHVC within $200$ kpc with a mass in the range between $M_{min}\approx 10^2~M_{\odot}$, and $M_{max}\approx 10^7~M_{\odot}$. The simulation is run until the total number of simulated clouds is equal to that of the total number of HVCs or CHVCs i.e., $N_{sim}=N_{total}$. Then, we estimate the total mass of the HVCs and CHVCs within $200$ kpc.

\begin{figure}
	\centering 
	\includegraphics[width=0.8\textwidth]{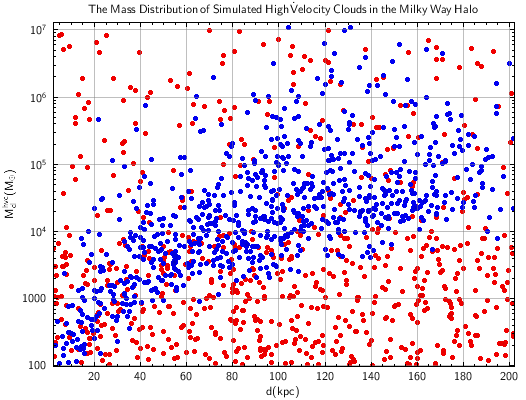}	
	\caption{The mass distribution of MC simulated HVCs is given as a function of the distance $d$ from the Earth, in the Milky Way halo within $200$ kpc. The blue dots represent the mass of the clouds that are already in the catalog. It is clearly seen that a significant number of smaller mass clouds are missing from the catalog. The total mass of the simulated HVCs is about $(7 \pm 2) \times 10^9 \, M_\odot$.}
  \label{fig:hvcmasssimulation}
\end{figure}

\begin{figure}
	\centering 
	\includegraphics[width=0.8\textwidth]{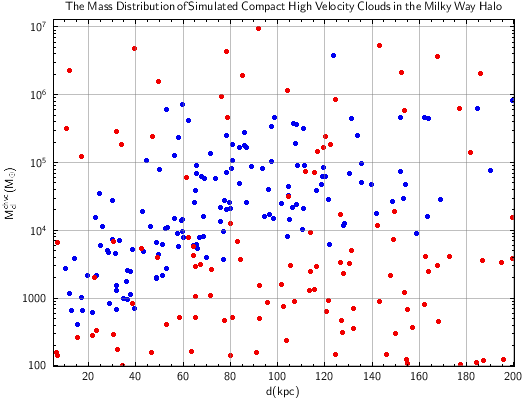}	
	\caption{Same as Fig. \ref{fig:hvcmasssimulation}, but for the case of CHVCs. The total mass of the simulated CHVCs is about $(5 \pm 1) \times \, 10^7 M_\odot$.}
  \label{fig:chvcmasssimulation}
\end{figure}

\begin{figure}
	\centering 
	\includegraphics[width=0.8\textwidth]{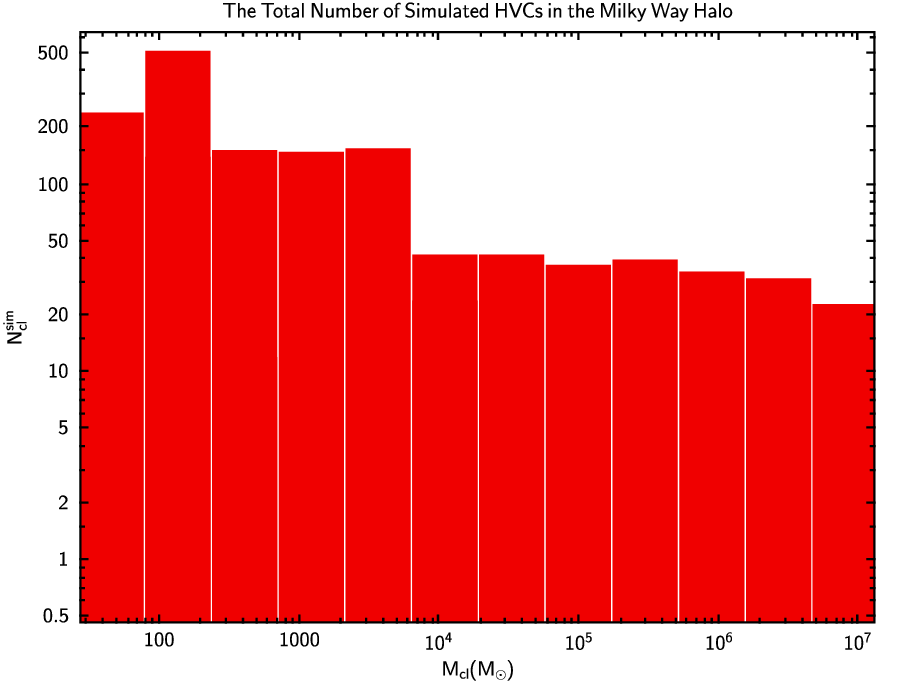}	
	\caption{The histogram shows the distribution in the mass of $1200$ simulated HVCs.}
  \label{fig:numbersimulatedhvc}
\end{figure}

\begin{figure}
	\centering 
	\includegraphics[width=0.8\textwidth]{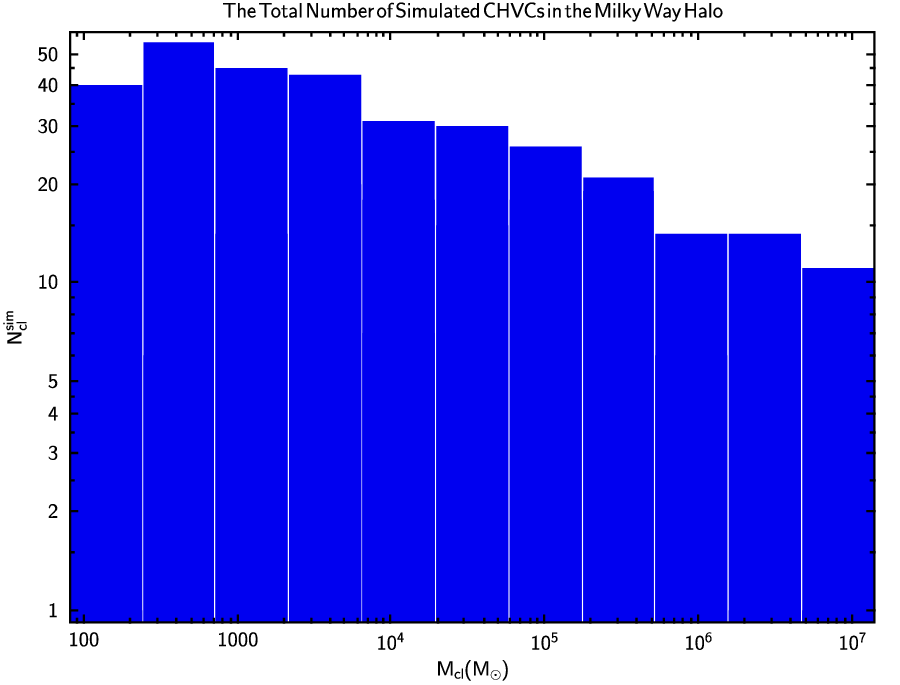}	
	\caption{Same as Fig. \ref{fig:numbersimulatedhvc}, but for the case of CHVCs. The number of the simulated CHVCs is $300$.}
  \label{fig:numbersimulatedchvc}
\end{figure}

The results of the simulation are shown in Figs. \ref{fig:hvcmasssimulation}--\ref{fig:numbersimulatedchvc}. In the case of HVCs we found that $152$ clouds with individual cloud mass between $10^2 M_\odot$ and $10^4 M_\odot$ might be missing in the observations. The simulation results show that the total mass of HVCs in the Milky Way halo should be $\sim (7 \pm 2) \times 10^9 M_\odot$, however, the catalogued clouds give the estimated total HVCs mass of $\sim (6 \pm 1) \times 10^9 M_\odot$. In the case of CHVCs we found that the observations might have missed about $100$ smaller mass CHVCs, and as a result, the total CHVCs mass of the simulated clouds is about $(5 \pm 1) \times 10^7 M_\odot$, while the total CHVCs mass of the cataloged clouds is $\sim (3 \pm 2) \times 10^7 M_\odot$. We can safely say that the total mass of the Milky Way halo in the form of HVCs and CHVCs should be $\sim (7 \pm 2) \times 10^9 M_\odot$.

We would like to caution that we have assumed that the HVCs and CHVCs distribution follows the NFW profile. There are other models that could explain the dark matter distribution in the galactic halos, e.g. the Einasto, Moore or Burkert density distribution. One can also use these distributions to perform MC simulation, but, the change in the estimates of the total mass is \textit{negligible}.

\section{Milky Way Halo Baryonic Mass Fraction in HVCs and CHVCs \label{baryonicfraction}}
We have estimated the total Milky Way halo mass in the form of HVCs and CHVCs. Our aim in this section is to estimate the contribution these clouds might give to the baryonic mass of the Milky Way halo. \citealp{flynn2006} estimated an upper limit on the baryonic mass of the Milky Way halo and showed that it is approximately $M_b^{MW} \simeq 6.10 \times 10^{10}~M_{\odot}$ up to $50$ kpc \citep{bland2016}. The baryonic mass is in the form of stars, dust, gas and ionized gas present in the Milky Way.

In the previous section, we estimated that the total Milky Way halo mass in the form of HVCs and CHVCs is $M_{cl}\approx (7 \pm 2) \times 10^{9} ~M_{\odot}$ up to $200$ kpc. It is seen that this mass is $\simeq 10^7~M_{\odot}$ up to $50$ kpc, which is much less than the baryonic mass of the Milky Way. One can easily estimate the baryonic mass fraction in the HVCs and CHVCs up to $50$ kpc as
\begin{equation}
MW_b^{c}= \frac{M_c^{corr}}{M_b^{MW}} \simeq \frac{10^7 M_{\odot}}{10^{10}~M_{\odot}} \simeq 0.001.
\end{equation} 
Hence, we can say that about $0.1 \%$ of the Milky Way baryonic mass is in the form of HVCs and CHVCs up to $50$ kpc. If we assume a case that the baryonic mass of the Milky Way halo up to $200$ kpc is about $10^{10}~M_{\odot}$, we can say that the Milky Way halo baryonic mass in the form of HVCs and CHVCs is $10 \%$. However, the above value can be considered as an upper limit on the fraction of the baryonic mass of the Milky Way halo. The obtained values of the mass of HVCs and CHVCs depend upon the estimation of the distances which rely on the assumption motivated by \citealp{olano2008}. Once the exact distances of the HVCs and CHVCs are observed one can estimate the exact baryonic mass fraction of the Milky Way halo in the form of HVCs and CHVCs. 
One should note that we have estimated the total HI mass of the HVCs and CHVCs in the Milky Way halo, indeed there is a fraction of ${\rm H_2}$ that has been detected in HVCs using Far Ultraviolet Spectroscopic Explorer (FUSE) observations of interstellar ${\rm H_2}$ \citep{richter2001}. The fraction of ${\rm H_2}$ is more dominant in the case of IVCs with $|v_{LSR}|$ of $20-90$ km/s \citep{tchernyshyov2022}. \citealp{tchernyshyov2022} have shown that the fraction of ${\rm H_2}$ is about $0.02$ in these clouds. If one can assume that the cataloged HVCs and CHVCs by \citealp{moss2013} have the same fraction of ${\rm H_2}$ the estimated total mass increases, but negligibly.

Relying on the estimated baryonic fraction of the Milky Way halo in the form of HVCs and CHVCs can one ask where the rest of the baryons in the halo reside. It might be the case that baryons are in the form of hot gas present in the galactic halos which is responsible to give rise to the rotational kinetic Sunyaev–Zeldovich (rkSZ) effect \citep{matilla2020}, and also they might be in the form of virial clouds present in the galactic halos \citep{de1995scenario, tahir2019a, tahir2019b, qadir2019, tahir2021}. The suggestion that the missing baryons might be present in these forms comes from the analysis of the cosmic microwave background (CMB) data of WMAP and {\it Planck} for various nearby edge-on spiral galaxies that showed the existence of a temperature asymmetry with respect to the considered galaxy's rotational axis, with the important characteristic of being almost frequency independent. This gave a strong indication of a Doppler shift effect induced by the galactic halo rotation \citep{de2011possible, de2014planck, de2015planck, gurzadyan2015planck, de2016triangulum, gurzadyan2018messier, de2019rotating, de1995observing}. Indeed, recently it was seen that the contribution of the rotational kinetic Sunyaev-Zeldovich (rkSZ) effect in the observed asymmetry towards M31 as seen by the {\it Planck} data is negligible \citep{tahir2022rksz}, making virial clouds a stronger candidate to contain a significant amount of effect in the galactic halo baryons. 

\section{Final Discussion and Summary \label{results}}
Galactic halos are far more mysterious and difficult to study than galactic disks. One thing from the flat rotational curves of the galaxies is sure that there is more matter outside the galactic disks, but the nature of this matter is still unknown. HVCs populate the galactic halos in large quantities but the exact distances of these clouds are still not estimated, since there are only a few early type stars in the halos to do spectroscopic measurements. 

We estimated the distances of HVCs and CHVCs observed by the Parkes telescope in the GASS survey and given in the catalog by \citealp{moss2013}. Following \citealp{olano2004, olano2008} we assumed that the cataloged HVCs and CHVCs are a part of the stream of clouds that are present in the Milky Way and they have the same kinematic properties as MS. These clouds were classified in various categories, as population MS, W, AC, depending upon their sky positions and radial velocities. We use two models the kinematic model and the dynamic model to estimate the distances of these clouds (see Section \ref{distance}). It was seen that for the case of population MS the estimated distance was in the range $20$ to $250$ kpc, for the case of population AC with sky position $0<l<180$ for $b>-15$ with radial velocity range from $-460$ km/s to $150$ km/s the estimated distance was in the range $5$ to $200$ kpc, for the case of population AC with sky position $100<l<180$ for $b<-15$ with radial velocity range from $-460$ km/s to $150$ km/s was in the range $1$ to $5$ kpc, and for the case of population W the estimated distance was in the range $5$ to $15$ kpc. The estimated distance of each HVC and CHVC population is in agreement with the distance limits estimated in Table-1 by \citealp{olano2008}.

Once the HVCs and CHVCs distances have been estimated, one is in the position to estimate the HI mass of each considered HVC and CHVC. We used eq. (\ref{mass}) and estimated the mass of a considered single HVC and CHVC (see Fig. \ref{fig2a}-\ref{fig2b}). We then estimated the missing HVCs and CHVCs in the observations by using MC simulations and obtained the estimates of their mass. The total mass of HVCs and CHVCs in the Milky Way halo turns out to be $\simeq (7 \pm 2) \times 10^{9} ~M_{\odot}$, which appears to be much less than the total baryonic mass of the Milky Way halo that is  $\simeq 6.1 \times 10^{10}~M_{\odot}$ up to 50 kpc \citep{flynn2006}. We, therefore, concluded that about $0.1 \%$ of the Milky Way halo baryon mass is in the form of the HVCs and CHVCs. 

Before closing the paper we would again like to caution that these estimates concerning the fraction of baryonic mass in the form of HVCs and CHVCs depend upon the estimates of the distances which are based on the assumption that all these HVCs and CHVCs come from the MCl. In order to better estimate the fraction of baryonic mass one needs to see the observed distances of these clouds. Our estimates just put an upper bound on the baryonic mass fraction of the Milky Way halo in the form of HVCs and CHVCs.

\section*{Acknowledgments}
NT would like to thank the Institute of Astrophysics of the Canary Islands, Tenerife, Spain for its collaboration in the project. NT would also like to acknowledge the hospitality and support of Mart\'{i}n L\'{o}pez Corredoira during the stay in Tenerife. FDP acknowledge the Euclid and TAsP projects of INFN. MLC's research was supported by the Chinese Academy of Sciences President’s International Fellowship Initiative grant number 2023VMB0001 and the grant PID2021-129031NB-I00 of the Spanish Ministry of Science (MICINN).

%\bibliographystyle{plain} 
%\bibliography{mybib} 

\begin{thebibliography}{000} %for 3 digits
%\begin{thebibliography}{00}  %for 2 digits
%\begin{thebibliography}{0}    %for 1 digit


\bibitem[Adams et al.(2013)]{adams2013} Adams, E., A., Giovanelli, R., \& Haynes, M., P. A. 2013, ApJ, 768, 77. 

\bibitem[Ade et al.(2016)]{ade2016} Ade, P., A., R., Aghanim, N., Arnaud, M., et al. 2016, A\&A, 594, A13.

\bibitem[Adelman-McCarthy et al.(2007)]{adelman2007} Adelman-McCarthy, J., K., Ag{\"u}eros, M., A., Allam, S., S. et al. 2007, ApJS, 172, 634. 

\bibitem[Bekki et al.(2007)]{bekki2007} Bekki, K., \& Chiba, M. 2007, MNRAS, 381, L16.

\bibitem[Beers et al.(1996)]{beers1996} Beers, T., C., Wilhelm, R., Doinidis, S., P. et al. 1996, ApJS, 103, 433.

\bibitem[Blitz et al.(1999)]{blitz1999} Blitz, L., Spergel, D., N., Teuben, P., J., et al. 1999, ApJ, 514, 818.

\bibitem[Bland-Hawthorn et al.(1998)]{bland1998} Bland-Hawthorn, J., Veilleux, S., Cecil, G., N., et al. 1998, MNRAS, 299, 611.

\bibitem[Bland-Hawthorn et al.(2001)]{bland2001} Bland-Hawthorn, J. \& Putman, M., E., 2001, ASP Conference Proceedings, 240, 369.

\bibitem[Bland-Hawthorn et al.(2002)]{bland2002} Bland-Hawthorn, J., \& Maloney, P., R. 2002,  ASP Conference Proceedings, 254, 267.

\bibitem[Bland-Hawthorn et al.(2016)]{bland2016} Bland-Hawthorn, J., \& Gerhard, O. 2016, ARA\&A, 54, 529.

\bibitem[Bonnar(1956)]{bonnor1956} Bonnor, W., B. 1956, MNRAS, 116, 351.

\bibitem[Braun et al.(1999)]{braun1999} Braun, R., \& Burton, W., B. 1999, A\&A, 341, 437.

\bibitem[Br{\"u}ns et al.(2001)]{bruns2001} Br{\"u}ns, C., Kerp, J., \& Pagels, A. 2001, A\&A, 370, L26.

\bibitem[Brown et al.(2004)]{brown2004} Brown, W., R., Geller, M., J., Kenyon, S., J. et al. 2004, ApJ, 127, 1555.

\bibitem[Burkert(2003)]{burkert2003} Burkert, A. 2003, ApSS, 284,697.

\bibitem[Cen et al.(1999)]{cen1999} Cen, R., \& Ostriker, J., P. 1999, ApJ, 514, 1.

\bibitem[Cen et al.(2006)]{cen2006} Cen, R., \& Ostriker, J., P. 2006, ApJ, 650, 560.

\bibitem[Connors et al.(2006)]{connors2006} Connors, T., W., Kawata, D., \& Gibson, B., K. 2006, MNRAS, 371, 108.

\bibitem[De Paolis et al.(1995a)]{depaolis1995a} De Paolis, F., Ingrosso, G., Jetzer, P., et al. 1995, A\&A, 295, 567.

\bibitem[De Palois et al.(1995b)]{de1995scenario} De Paolis, F., Ingrosso, G., Jetzer, P., et al. 1995, A\&A, 295, 567.

\bibitem[De Paolis et al.(1995c)]{de1995observing} De Paolis, F., Ingrosso, G., Jetzer, P., et al. 1995, A\&A, 299, 647.

\bibitem[De Paolis et al.(2011)]{de2011possible} De Paolis, F., Gurzadyan, V., Ingrosso, G., et al. 2011, A\&A, 534, L8.

\bibitem[De Paolis et al.(2014)]{de2014planck} De Paolis, F., Gurzadyan, V., Nucita, A., A., et al. 2014, A\&A, 565, L3.
	
\bibitem[De Palois et al.(2015)]{de2015planck} De Paolis, F., Gurzadyan, V., Nucita, A., A., et al. 2015, A\&A, 580, L8.

\bibitem[De Paolis et al.(2019)]{de2019rotating} De Paolis, F., Gurzadyan, A., Nucita, A., A., et al. 2019, A\&A, 629, A87.

\bibitem[De Paolis et al.(2016)]{de2016triangulum} De Paolis, F., Gurzadyan, V., Nucita, A., A., et al. 2016, A\&A, 593, A57.

\bibitem[Ebert(1955)]{ebert1955} Ebert, R. 1995, ZAP, 37, 217.

\bibitem[Faridani et al.(2014)]{faridani2014} Faridani, S., Fl{\"o}er, L., {Kerp}, J. et al. 2014, A\&A, 563, A99.

\bibitem[Flynn et al.(2006)]{flynn2006} Flynn, C., Holmberg, J., Portinari, L., et al. 2006, MNRAS, 372, 1149.

\bibitem[Fraser-McKelvie et al.(2011)]{fraser2011} Fraser-McKelvie, A., Pimbblet K., A., \& Lazendic, J., S. 2011, MNRAS, 415, 1961.

\bibitem[Fukugita et al.(1996)]{fukugita1996} Fukugita, M., Ichikawa, T., Gunn, J., E. et al. 1996, ApJ, 111, 1748.

\bibitem[Gardiner et al.(1996)]{gardiner1996} Gardiner, L., T., \& Noguchi, M. 1996, MNRAS, 278, 191.

\bibitem[Gerhard et al.(1996)]{gerhard1996} Gerhard, O., \& Silk, J. 1996, ApJ, 472, 34.

\bibitem[Giovanelli(1981)]{giovanelli1981} Giovanelli, R. 1981, Astron. J., 86, 1468.

\bibitem[Gurzadyan et al.(2015)]{gurzadyan2015planck} Gurzadyan, V., De Paolis, F., Nucita, A., A., et al. 2015, A\&A, 582, A77. 

\bibitem[Gurzadyan et al.(2018)]{gurzadyan2018messier} Gurzadyan, V., De Paolis, F., Nucita, A., A., et al. 2018, A\&A, 609, A131.

\bibitem[Gunn et al.(1996)]{kathine1996} Gunn, K., F., \& Thomas, P., A. 1996, MNRAS, 281, 1133.

\bibitem[Gunn et al.(1998)]{gunn1998} Gunn, J., E., Carr, M., Rockosi, C. et al. 1998, ApJ, 116, 3040.

\bibitem[Gunn et al.(2006)]{gunn2006} Gunn, J., E., Siegmund, W., A., Mannery, E., J. et al. 2006, ApJ, 131, 2332.

\bibitem[Hopp et al.(2007)]{hopp2007} Hopp, U., Schulte-Ladbeck, R., E., \& Kerp, J. 2007, MNRAS, 374, 1164.

\bibitem[Hulsbosch(1968)]{hulsbosch1968} Hulsbosch, A., N., M. 1968, BAIN, 20, 33.

\bibitem[Haud(1990)]{haud1990} Haud, U. 1990, A\&A, 230, 145. 

\bibitem[Hulsbosch et al.(1973)]{hulsbosch1973} Hulsbosch, A., N., M., \& Oort, J., H. 1973, A\&A, 22, 153.

\bibitem[Hsu et al.(2011)]{hsu2011} Hsu, W., H., Putman, M., E., Heitsch, F., et al. 2011, ApJ, 141, 57.

\bibitem[Kroupa(2001)]{kroupa2001} Kroupa, P. 2001, MNRAS, 322, 231.

\bibitem[Lehner et al.(2022)]{lehner2022} Lehner, N., Howk, J., C., Marasco, A., et al. 2022, MNRAS, 513, 3228.

\bibitem[Li et al.(2018)]{li2018a} Li, J., T., Bregman, J., N., Wang, Q., D., et al. 2018, ApJL, 855, L24.

\bibitem[L{\'o}pez-Corredoira et al.(1999)]{martin1999}  L{\'o}pez-Corredoira, M., Beckman, J., E.,  \& Casuso, E. 1999, A\&A, 351, 920.

\bibitem[Lucchini et al.(2021)]{lucchini2021} Lucchini, S., D’Onghia, E., \& Fox, A., J. 2021, ApJ, 921, L367.

\bibitem[Mathewson et al.(1974)]{mathewson1974magellanic} Mathewson, D., S., Cleary, M., N., \& Murray, J., D. 1974, In Symposium-IAU, 60, 617.

\bibitem[Matilla et al.(2020)]{matilla2020} Matilla, J., M., Z., \& Haiman, Z. 2020, PRD, 101, 083016.

\bibitem[Maller et al.(2004)]{maller2004} Maller, A., H., \& Bullock, J., S. 2004, MNRAS, 355, 694.

\bibitem[Mastropietro et al.(2005)]{mastropietro2005} Mastropietro, C., Moore, B., Mayer, L., et al. 2005, MNRAS, 363, 509.

\bibitem[Meurer et al.(1985)]{meurer1985} Meurer, G., R., Bicknell, G., V., \& Gingold, R., A. 1985, Proc. Astron. Soc. Aust., 6, 195.

\bibitem[Mirabel(1981)]{mirabel1981} Mirabel, I., F. 1981, ApJ, 250, 528.

\bibitem[Muller et al.(1963)]{muller1963} Muller, C., A., Oort, J., H., \& Raimond, E.  1963 , C. R. Acad. Sci. III, 257, 1661.

\bibitem[Moore et al.(1994)]{moore1994} Moore, B., and Davis, M. 1994, MNRAS, 270, 209.

\bibitem[Moss et al.(2013)]{moss2013} Moss, V., A., McClure-Griffiths, N., M., Murphy, T., et al. 2013, ApJS, 209, 12.

\bibitem[Navarro et al.(1996)]{navarro1996} Navarro, J., F., Frenk, C., S., \& White, S., D., M. 1996, ApJ, 462, 563.

\bibitem[Nesti et al.(2013)]{nesti2013} Nesti, F., \& Paolo S. 2013, JCAP, 2013, 16.

\bibitem[Nicastro et al.(2008)]{nicastro2008} Nicastro, F., Mathur, S.,  \& Elvis, M. 2008, Sci., 319, 55.

\bibitem[Olano(2004)]{olano2004} Olano, C., A. 2004, A\&A, 423, 895.

\bibitem[Olano(2008)]{olano2008} Olano, C., A. 2008, A\&A, 485, 457.

\bibitem[Oort(1996)]{oort1966} Oort, J., H. 1996, BAIN, 18, 421.

\bibitem[Peek et al.(2007)]{peek2007} Peek, J., E., G., Putman, M., E., McKee, C., F., et al. 2007, ApJ, 656, 907.

\bibitem[Pisano et al.(2004)]{pisano2004} Pisano, D., J., Barnes, D., G., Gibson, B. K., et al. 2004, ApJ, 610, L17.

\bibitem[Pier et al.(2003)]{pier2003} Pier, J., R., Munn, J., A., Hindsley, R., B. et al. 2003, ApJ, 125, 1559.

\bibitem[Putman et al.(2003)]{putman2003} Putman, M., E., Bland-Hawthorn, J., Veilleux, S., et al. 2003, ApJ, 597, 948.

\bibitem[Putman et al.(2012)]{putman2012} Putman, M., E., Peek, J., E., G., \& Joung, M., R. 2012,  ARA\&A, 50, 491.

\bibitem[Qadir et al.(2019)]{qadir2019} Qadir, A., Tahir, N., \& Sakhi, M. 2019, PRD, 100, 043028.

\bibitem[Richter et al.(2001)]{richter2001} Richter, P., Sembach, K., R., Wakker, B., P. et al. 2001, ApJL, 562, L181.

\bibitem[Richards et al.(2018)]{richads2018} Richards, E., E., van Zee, L., Barnes, K., L., et al. 2018, MNRAS, 476, 5127.

\bibitem[Salpeter(1955)]{salpeter1955} Salpeter, E., E. 1955, ApJ, 121, 161.

\bibitem[Siegel et al.(2005)]{siegel2005} Siegel, M. H., Majewski, S. R., Gallart, C., et al. 2005, ApJ, 623, 181.

\bibitem[Simon et al.(2002)]{simon2002} Simon, J., D.,  \& Blitz, L. 2002, ApJ, 574, 726.

\bibitem[Smoker et al.(2011)]{smoker2011} Smoker, J., V., Fox, A., J., \& Keenan, F., P. 2011, MNRAS, 415, 1105.

\bibitem[Sommer-Larsen(2006)]{sommer2006} Sommer-Larsen, J. 2006, ApJL, 644, L1.

\bibitem[Stoughton et al.(2002)]{stoughton2002} Stoughton, C., Lupton, R., H., Bernardi, M. et al. 2002, ApJ, 123, 485.

\bibitem[Tahir et al.(2019a)]{tahir2019a} Tahir, N., De Paolis, F., Qadir, A., et al. 2019, IJMPD, 28, 1950088.

\bibitem[Tahir et al.(2019b)]{tahir2019b}  Tahir, N., De Paolis, F., Qadir, A., et al. 2019, AJOM, 8, 193.

\bibitem[Tahir et al.(2022)]{tahir2022rksz} Tahir, N., Qadir, A., De Paolis, F., et al. 2022, A\&A, 664, A30.

\bibitem[Tahir et al.(2021)]{tahir2021} Tahir, N., Qadir, A., Sakhi, M. et al. 2021, EPJC, 81, 827.

\bibitem[Tchernyshyov(2022)]{tchernyshyov2022} Tchernyshyov, K. 2022, ApJ, 931, 78.

\bibitem[Thom et al.(2006)]{thom2006} Thom, C., Putman, M., E., Gibson, B., K., et al. 2006, ApJ, 638, L97.

\bibitem[van Woerden et al.(2000)]{woerden2000} van Woerden, H., Wakker, B., P., Peletier, R., F., et al. 2000,  ASPCS, 218, 407. 

\bibitem[Wakker et al.(1997)]{wakker1997} Wakker, B., P. \& van Woerden, H. 1997, ARA\&A, 35, 217.

\bibitem[Wakker(2001)]{wakker2001} Wakker, B., P. 2001, ApJS, 136, 463.

\bibitem[Wakker et al.(2001)]{wakker2001a} Wakker, B., P., Kalberla, P., M., W., Van Woerden, H., et al. 2001, ApJS, 136, 537.

\bibitem[Wakker et al.(2007)]{wakker2007a} Wakker, B., P., York, D., G., Howk, J., C., et al. 2007, ApJ, 670, L113.

\bibitem[Westmeier(2017)]{westmeier2017}  Westmeier, T. 2018, MNRAS, 474, 289.

\bibitem[Yoshizawa et al.(2003)]{yoshizawa2003} Yoshizawa, A., M., \& Noguchi, N. 2003, MNRAS, 339, 1135.

\bibitem[York et al.(2000)]{york2000} York, D., G., Adelman, J., Anderson, J., E., Jr. et al. 2000, ApJ, 120, 1579.

\bibitem[Zaritsky et al.(1989)]{zaritsky1989} Zaritsky, D., Olszewski, E., W., Schommer, R., A., et al. 1989, ApJ, 345, 759.

\bibitem[Zwaan(2001)]{zawaan2001} Zwaan, M., A. 2001, MNRAS, 325, 1142.

\bibitem[Zhang et al.(2021)]{zhang2021} Zhang, Yi, Liu, R., Y., Shi Shao, H., L., et al. 2021, ApJ, 911, 58.

\end{thebibliography}

\end{document}